\documentclass[aps,prd,reprint,superscriptaddress,nofootinbib]{revtex4-2}
\usepackage{amsmath,amssymb,bm,mathrsfs}
\usepackage[sort&compress]{natbib}
\usepackage{srcltx}
\usepackage{dsfont}
\usepackage{epsfig}
\usepackage{slashed}
\usepackage{bbold}
\usepackage{psfrag}
\usepackage[dvipsnames]{xcolor}
\PassOptionsToPackage{caption=false}{subfig}
\usepackage{subfig}
\usepackage{xfrac}
\usepackage{multirow}
\usepackage{booktabs}
\usepackage{graphicx}
\usepackage{soul}
\usepackage[colorlinks=true
,urlcolor=blue
,anchorcolor=blue
,citecolor=blue
,filecolor=blue
,linkcolor=red
,menucolor=blue
,linktocpage=true
,pdfproducer=medialab
,pdfa=true
]{hyperref}
\usepackage{cleveref}
\usepackage{setspace}
\usepackage{relsize}
\usepackage{tikz,tikz-feynman,enumerate,comment}
\usepackage{physics}
\usepackage[bottom]{footmisc}
\usepackage{array} 
\usepackage{pgfplots}

\makeatletter
\g@addto@macro\bfseries{\boldmath}
\let\Hy@backout\@gobble
\makeatother

\usepackage{xspace}
\newcommand{\FCCee}{\mbox{FCC-$ee$}\xspace}
\newcommand{\fccee}{\FCCee}

\tikzset{
    vector/.style={decorate, decoration={snake}, draw},
    fermion/.style={postaction={decorate},
        decoration={markings,mark=at position .55 with {\arrow{>}}}},
    fermionbar/.style={draw, postaction={decorate},
        decoration={markings,mark=at position .55 with {\arrow{<}}}},
    fermionnoarrow/.style={},
    gluon/.style={decorate,
        decoration={coil,amplitude=4pt, segment length=5pt}},
    scalar/.style={dashed, postaction={decorate},
        decoration={markings,mark=at position .55 with {\arrow{>}}}},
    scalarbar/.style={dashed, postaction={decorate},
        decoration={markings,mark=at position .55 with {\arrow{<}}}},
    scalarnoarrow/.style={dashed,draw},
	vectorscalar/.style={loosely dotted,draw=black, postaction={decorate}},
}

\usetikzlibrary{decorations.pathreplacing,calligraphy}

\usepackage[normalem]{ulem}
\usepackage[caption=false]{subfig}

\newcommand{\slepR}{\widetilde{E}}
\newcommand{\slepL}{\widetilde{L}}
\newcommand{\Bino}{\widetilde{B}}
\newcommand{\Wino}{\widetilde{W}}

\definecolor{MyRed}{HTML}{a0586b}
\definecolor{MyYellow}{HTML}{dea04c}
\definecolor{MyPurple}{HTML}{946e96}
\definecolor{MyGreen}{HTML}{33cd32}
\definecolor{MyBlue}{HTML}{6394ea}
\definecolor{MyGrey}{HTML}{717171}

\newcommand{\bfRed}[1]{{\textbf{\textcolor{MyRed}{#1}}}}
\newcommand{\bfYellow}[1]{{\textbf{\textcolor{MyYellow}{#1}}}}
\newcommand{\bfPurple}[1]{{\textbf{\textcolor{MyPurple}{#1}}}}
\newcommand{\bfBlue}[1]{{\textbf{\textcolor{MyBlue}{#1}}}}
\newcommand{\bfGreen}[1]{{\textbf{\textcolor{MyGreen}{#1}}}}
\newcommand{\bfGrey}[1]{{\textbf{\textcolor{MyGrey}{#1}}}}

\graphicspath{{figs/}}

\begin{document}

\preprint{XXX}

\title{
Imprints of supersymmetry at a future $Z$ factory}

\author{Simon Knapen}
\affiliation{Theory Group, Lawrence Berkeley National Laboratory, Berkeley, CA 94720, USA}
\affiliation{Berkeley Center for Theoretical Physics, University of California, Berkeley, CA 94720, USA}

\author{Kevin Langhoff}

\affiliation{Theory Group, Lawrence Berkeley National Laboratory, Berkeley, CA 94720, USA}
\affiliation{Berkeley Center for Theoretical Physics, University of California, Berkeley, CA 94720, USA}

\author{Zoltan Ligeti}
\affiliation{Theory Group, Lawrence Berkeley National Laboratory, Berkeley, CA 94720, USA}
\affiliation{Berkeley Center for Theoretical Physics, University of California, Berkeley, CA 94720, USA}

\begin{abstract}

We study the discovery potential of $Z$ branching ratios due to contributions induced by the MSSM electroweak sector, assuming that the squarks and gluinos are heavy. Precision measurements at a future $Z$ factory would yield sensitivity to MSSM that is complementary to direct searches at the LHC, provided that the systematic uncertainties can be reduced to a level comparable to the expected statistical uncertainties. 
\end{abstract}

\maketitle


\paragraph*{\bf Introduction} A future $Z$ factory such as the \FCCee or CEPC present opportunities to measure the electroweak sector of Standard Model (SM) with unprecedented precision. The \FCCee is anticipated to produce around $5\times 10^{12}$ $Z$ bosons, about $10^5$ times more than LEP~\cite{FCC:2018byv, Benedikt:2653669}. 
This increase in statistics would greatly enhance the sensitivity to loop-induced corrections to $Z$-pole observables. These effects can be studied in a relatively model-independent manner using the SMEFT \cite{Buchmuller:1985jz,Grzadkowski:2010es,Isidori:2023pyp,Brivio:2017vri} and performing global fits \cite{DeBlas:2019qco,deBlas:2019rxi,deBlas:2022ofj,Celada:2024mcf}. 
In this Letter we take the complementary model dependent approach and quantify the contributions of the minimal supersymmetric standard model (MSSM) electroweak sector to the $Z$ branching ratios. 
This calculation yields a new estimate of a $Z$-factory's sensitivity to supersymmetry, when combined with an estimate of the achievable uncertainties on the branching ratio measurements. 

Previously in the LEP/SLD era, electroweak precision tests were identified as a promising method to search for effects of supersymmetry. Often, this focused on oblique/universal corrections, i.e., corrections to the gauge boson vacuum polarizations~\cite{Peskin:1990zt,Peskin:1991sw,Drees:1991zk}. 
It was also observed that the MSSM gives rise to vertex corrections which may lead to interesting flavor universal, flavor non-universal, or flavor violating decays \cite{Hagiwara:1990st,Boulware:1991vp,Illana:2002tg}. 
Further, the sensitivities of \FCCee and CEPC have been studied in the context of natural SUSY~\cite{Fan:2014vta,Fan:2014axa,Essig:2017zwe,Liu:2017msv}. The latter studies primarily focused on the colored sector of the MSSM and its impact on the oblique corrections, as well as the 
$Z\to b\bar b$, $h \to gg$ and $h\to\gamma\gamma$ branching ratios. 
We extend this work by considering the impact of the electroweak sector on the $Z\to \ell\ell$ branching ratios.

In the context of supersymmetry, a SM-like Higgs with mass of 125 GeV implies either a rather non-minimal model or colored sparticles with masses $\gtrsim 5$ TeV (see, e.g.,~\cite{PardoVega:2015eno}).
Such heavy colored sparticles are well outside the reach of the HL-LHC and a future \mbox{$Z$-factory}, as we shall see below.
There are no such theoretical lower bounds for the electroweak states however, which in some cases will remain unconstrained at the HL-LHC, even with masses well below a TeV.
More broadly, studying (simplified) models is important to evaluate the discovery potential of the \fccee and CEPC: 
In a concrete model, operators and observables can be correlated in a manner which is not captured in an effective field theory analysis.  

\begin{figure*}[t!]
\includegraphics[width=\columnwidth]{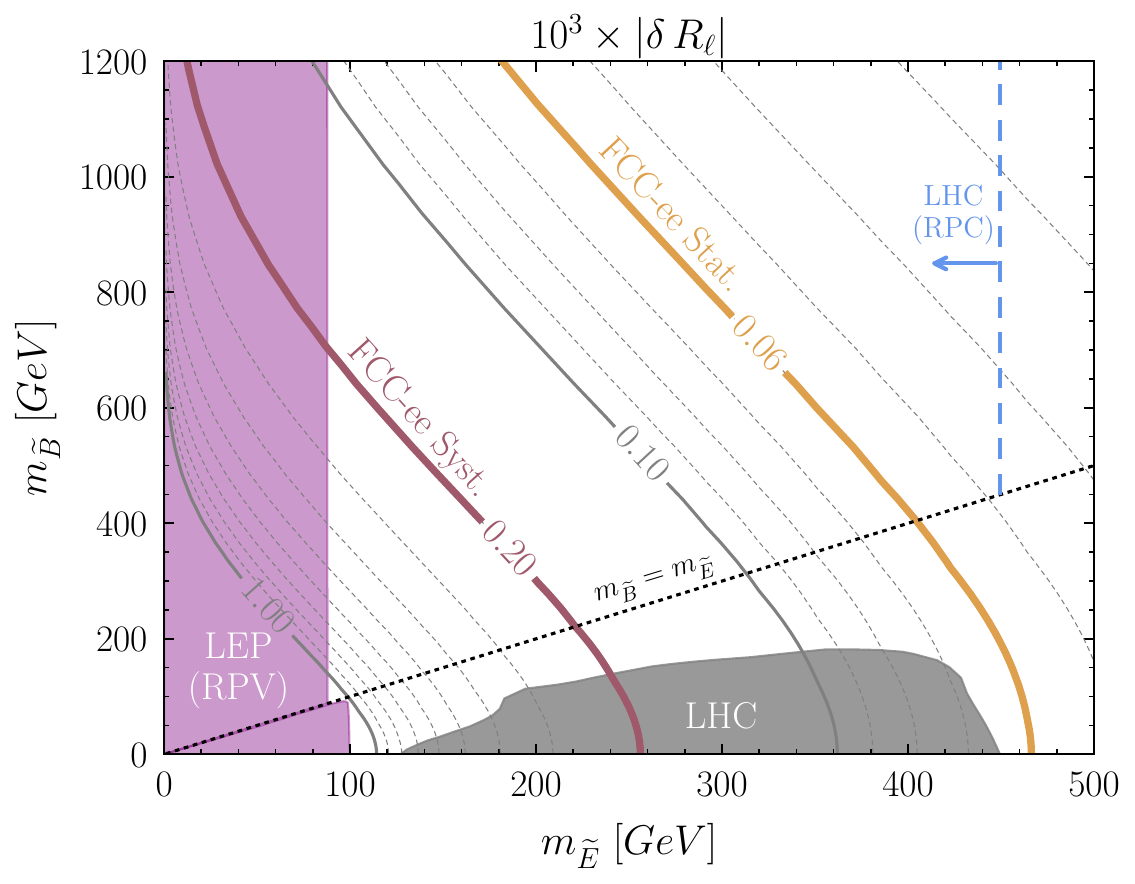}
\hfill
\includegraphics[width=\columnwidth]{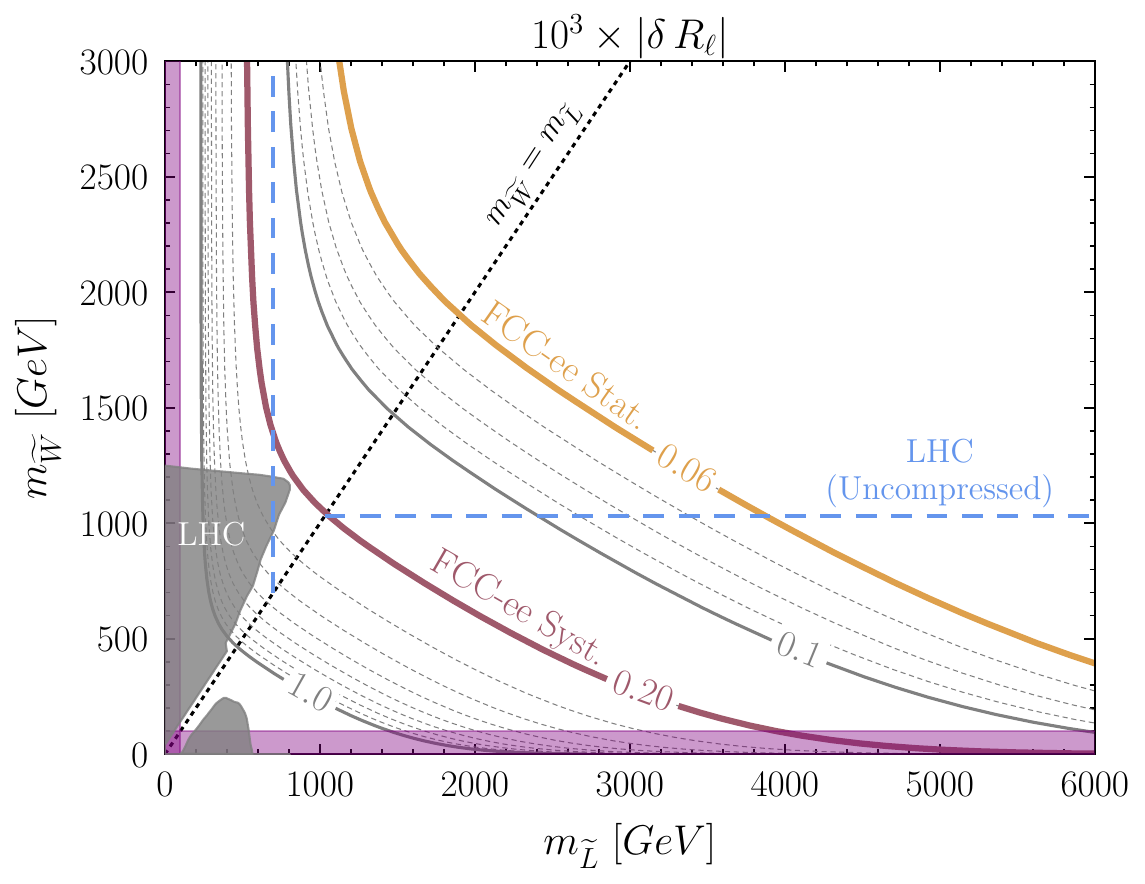}
    \caption{\textbf{Left:} $|\delta R_\ell|$ for pure bino and a right-handed selectron simplified model. On the \bfYellow{yellow} (\bfRed{red}) contour, $|\delta R_\ell|$ matches the anticipated \fccee 1$\sigma$ statistical (systematic) uncertainty~\cite{FCC:2018byv}. We also show the current limit from the ATLAS $\slepR \to \ell +$MET search \cite{ATLAS:2019lff} (\bfGrey{gray}), the LEP $\slepR \to \ell +$MET bound (\bfPurple{purple}, below dotted line) \cite{DELPHI:2003uqw} and LEP RPV squark limit, as a proxy for the $ \slepR \to jj$ R-parity violating decay (\bfPurple{purple}, above dotted line) \cite{L3:2001xyr}. For $m_{\Bino}>m_{\slepR}$, the bounds are significantly stronger we assume R-parity  conservation (RPC) by allowing the $\slepR$ to decay to a lepton and a nearly massless gravitino (dashed \bfBlue{blue} line). 
    \textbf{Right:} Same, for pure wino and a left-handed selectron simplified model. The LHC bounds are reinterpretations of the $\ell^+\ell^-\,+$~MET bound (\bfGrey{gray}, below dotted line) \cite{ATLAS:2019lff} and the $\ell^\pm\ell^\pm\,+$~MET search \cite{CMS:2021cox} (\bfGrey{gray}, above dotted line) (see text for details). The LEP bounds (\bfPurple{purple}) were taken from \cite{ALEPH:2002gap,DELPHI:2003uqw}. The LHC bounds from direct production assuming an uncompressed spectrum, e.g., a nearly massless gravitino or bino LSP, are shown as a dashed \bfBlue{blue} line.
    }
\label{fig:delta Rl Bino}
\end{figure*}

\medskip
\paragraph*{\bf Setup and results} In this Letter we focus on 
\begin{align}
    R_\ell \equiv \frac{\Gamma(Z \rightarrow {\rm hadrons})}{\Gamma(Z \rightarrow \ell \bar{\ell})} \,,
\end{align}
with $\ell$ representing one of the SM leptons.
We consider corrections to $R_\ell$ in simplified models which are meant to represent specific sectors of the MSSM. We assume that only a single sfermion species ($\tilde{f}$) and a single gaugino $(\tilde{\chi})$ dominate the corrections to $R_\ell$. All sfermion mass matrices are assumed to be proportional to the identity matrix.
Our simplified models and the corresponding scaling of their contributions to $R_\ell$ are summarized in Table \ref{tab:models}; their full definitions and calculations are presented in Appendix~\ref{App:MSSM} and Appendix~\ref{App:Details of SUSY Corrections to $Z$ Decay Rate} respectively.  
Given that the correction to $R_\ell$ decreases for larger masses, we pick the sfermion in the pairs $(\tilde{f},\, \tilde{\chi})$ for a given $\tilde{\chi}$ by choosing the lightest sfermion which couples to $\tilde{\chi}$ under the assumption $m_{\slepR}< m_{\slepL}< m_{\tilde{q}}$. 
Here $\slepR~(\slepL)$ is the right (left) handed slepton and $\tilde{q}$ are the left and right handed squarks, which we take to be degenerate. 
The hierarchy of the sfermion masses is motivated by the fact that sfermion masses generically increase under renormalization group flow, proportional to the couplings of the gauge symmetries through which they interact. 

\begin{table}[b]
    \def\arraystretch{1.4}\tabcolsep 10pt
    \begin{tabular}{ll}
    \hline\hline
       $(\tilde{f},\, \tilde{\chi})$  & $\mathcal{O}\left(|\delta R_\ell|/R_\ell \right)$  \\
       \midrule
       $(\slepR,\, \Bino)$  & $(\alpha'/4\pi)\, (m_Z^2/m^2_{\slepR, \Bino})$\\
       $( \slepL,\, \Wino)$  & $(\alpha_W/4\pi)\, (m_Z^2/m^2_{\slepL, \Wino})$\\
       $(\tilde{q},\, \tilde{g})$  & $(\alpha_s/4\pi)\, (m_Z^2/m^2_{\tilde{q},\tilde{g}})$ \\
    \hline\hline
    \end{tabular}
    \caption{The simplified models considered and the parametric size of their contributions to $R_\ell$.}
    \label{tab:models}
\end{table}

Figure~\ref{fig:delta Rl Bino} shows $|\delta R_\ell|$ for the simplified models with light electroweakinos. 
The highlighted contours show the \FCCee's anticipated statistical uncertainty, $0.06 \times 10^{-3}$ (yellow), and the estimated systematic uncertainty, $0.2 \times 10^{-3}$ (red)~\cite{Bernardi:2022hny}. 
The sensitivity does not fully disappear if the Wino is decoupled, due to the oblique corrections induced by the sleptons (See Appendix~\ref{App:Details of SUSY Corrections to $Z$ Decay Rate}). This can be compared with global electroweak fits for the oblique parameters \cite{EuropeanStrategyforParticlePhysicsPreparatoryGroup:2019qin}, which yield a 2$\sigma$ sensitivity of $m_{\slepL}\gtrsim 640$~GeV. 
As expected, this is a stronger bound than considering only $R_\ell$ in the $m_{\Wino}\gg m_{\slepL}$ limit. We see however that the non-universal vertex correction to $R_\ell$ provides sensitivity beyond the oblique parameters for $m_{\Wino}\lesssim m_{\slepL}$.
The case of squarks and gluinos is discussed in Appendix~\ref{App:G+Q}, since the LHC constraints in this case are significantly stronger (see Figure~\ref{fig:g+q} therein). 

\medskip
\paragraph*{\bf LHC bounds and projections}
The LHC bounds are model dependent, because additional assumptions about the SUSY particle spectra\footnote{The $m_{\slepR}$, $m_{\slepL}$, $m_{\Wino}$ and $m_{\Bino}$ refer to the soft masses, which are the most relevant for the radiative corrections to $R_\ell$. The LHC bounds on the other hand depend on the physical masses, which can differ slightly from the soft masses due to electroweak symmetry breaking effects. For simplicity we neglect these differences in our discussions of the LHC bounds, since they are smaller than our current uncertainty on what the ultimate HL-LHC constraints will be.} and couplings are needed. We consider three scenarios:
\begin{enumerate}
    \item R-parity is conserved and the lightest supersymmetric particle (LSP) (bino or gravitino) is light enough such that phase space for the decay of the next-to-lightest supersymmetric particle (NLSP) is not significantly compressed. In this case we find that the direct searches at the \mbox{(HL-)LHC} typically outperform a precision measurement of $R_\ell$ at the \fccee or CEPC.
    \item If R-parity is conserved, but the bino LSP is close in mass to the NLSP ($\Wino$, $\slepL$, or $ \slepR$), then the $R_\ell$ measurement can be competitive with (HL-)LHC direct searches in part of the parameter space, as shown in Fig.~\ref{fig:delta Rl Bino}.
    \item R-parity is violated through operators causing the LSP to decay to first and second generations jets. In this case there are no competitive LHC limits on electroweak states and the $R_\ell$ constraint may be a major improvement upon the LEP bounds.
\end{enumerate}
Below we explain these qualitative points in more detail. 

For the $(\slepR,\, \Bino)$ model the $\slepR \to \ell\, +\,$MET search~\cite{ATLAS:2019lff} applies if $m_{\slepR}>m_{\Bino}\,+\,m_\ell$ (\bfGrey{gray} shading in Fig.~\ref{fig:delta Rl Bino}).
Given that the HL-LHC is expected to add a few hundred GeV in mass reach, we see that the \fccee/CEPC only has sensitivity in the compressed region $m_{\slepR}\approx m_{\Bino}$. This is complemented by the limited but robust bounds from LEP \cite{DELPHI:2003uqw}.

If $m_{\slepR}<m_{\Bino}+m_\ell$, the $\slepR$ could decay to $\ell$ + gravitino. 
This decay may be prompt or displaced, depending on the gravitino mass. 
For a prompt decay, the current bound is $m_{\slepR}\gtrsim 450$\,GeV~\cite{ATLAS:2019lff}, which already exceeds that from a future $R_\ell$ measurement at the \fccee/CEPC. 
This bound can be evaded by allowing the $\slepR$ to decay to a pair of jets, as can be realized in models of dynamical R-parity breaking \cite{Csaki:2013jza}.
This case is unconstrained at the LHC due to the low slepton production cross section relative to that of the dijet background. 
While LEP did not search for $\slepR\to jj$, we can use the squark search in the same channel as an approximate bound $m_{\slepR }\gtrsim 87$\,GeV~\cite{L3:2001xyr} (\bfPurple{purple} shading above dotted line in left-hand panel of Fig.~\ref{fig:delta Rl Bino}).

The $(\slepL,\, \Wino)$ model with a light bino or gravitino LSP is similarly subject to strong LHC bounds. For \mbox{$m_{\slepL}<m_{\Wino}$} ATLAS already bounded \mbox{$m_{\slepL}\gtrsim 700$ GeV} from direct slepton production \cite{ATLAS:2019lff,CMS:2020bfa}.
Similarly, for $m_{\slepL}>m_{\Wino}$ searches for direct chargino production already require $m_{\Wino}\gtrsim 1025$ GeV \cite{ATLAS:2024qxh}.
Bearing in mind that these bounds will continue to improve with the HL-LHC, they are likely to eliminate most of the parameter space that can be probed by the \fccee/CEPC, even with optimistic assumptions for the systematic uncertainties.

At the same time, these bounds are relaxed significantly if the bino LSP is compressed towards mass to the $\Wino$ or $\slepL$ NLSP \cite{ATLAS:2017vat}.\footnote{For $m_{\slepL} < m_{\Wino}$, the sneutrino ($\tilde \nu$) can be the LSP, with very similar phenomenology \cite{Katz:2009qx}.} 
In this case, there are direct bounds of $m_{\Wino,\slepL}\gtrsim 100$ GeV from LEP \cite{ALEPH:2002gap,DELPHI:2003uqw} (\bfPurple{purple} in Fig.~\ref{fig:delta Rl Bino}), in addition to LHC bounds from the $\Wino\to \slepL\to \Bino$ and $\slepL\to \Wino\to \Bino$ cascade decays, depending on the mass ordering. The most striking signatures are 
\begin{enumerate}
  \setlength{\itemsep}{0pt}
  \setlength{\parskip}{0pt}
    \item[\emph{i)}] three or more leptons + MET,
    \item[\emph{ii)}] two same sign leptons ($\ell^\pm\ell^\pm$) + MET,
    \item[\emph{iii)}] two opposite sign leptons ($\ell^+\ell^-$) + MET.
\end{enumerate}
ATLAS and CMS have carried out searches for the \mbox{$\Wino\to \slepL\to \Bino$} cascade using the three lepton and same sign lepton signatures \cite{CMS:2021cox,ATLAS:2018ojr}. 
For moderately compressed spectra they found the same sign dilepton channel to be the most constraining. The search in \cite{CMS:2021cox} assumed a 100\% branching ratio of the neutral wino component to the charged slepton + lepton final state. 
We rescaled the cross section with a factor of 1/2 to account for the branching ratio to the sneutrino + neutrino final state. The resulting bound is shown above the dotted line in the right-hand panel of Figure~\ref{fig:delta Rl Bino}.
The $\slepL\to \Wino\to \Bino$ cascade was not searched for by ALTAS and CMS, however we can reinterpret the \mbox{$\ell^\pm\ell^\pm\, +$ MET} \cite{CMS:2021cox} and \mbox{$\ell^+\ell^-\, +$ MET} searches \cite{ATLAS:2019lff,CMS:2020bfa}, as explained in Appendix~\ref{app:LHClimits}. 
We find that both limits are comparable, with the latter currently being slightly stronger. We therefore show our reinterpretation of the \mbox{$\ell^+\ell^-\, +$ MET} limit below the dashed line in the right-hand panel of Figure~\ref{fig:delta Rl Bino}. 
The limits described here will likely strengthen by at least 200 to 300 GeV with the HL-LHC data. This will likely eliminate most of the available parameter space if $m_{\Wino}>m_{\slepL}$, though for $m_{\Wino}<m_{\slepL}$ the \FCCee/CEPC may have complementary sensitivity. 

Finally, all current and likely future (HL-)LHC bounds are evaded if a $\slepL$ or $\Wino$ LSP decays to first and second generation quarks due to an R-parity violating interaction. (If the RPV couplings are such that the final state contains a top quark, ATLAS obtained $m_{\Wino}\gtrsim 360$ GeV~\cite{ATLAS:2021fbt}.)

\medskip

\paragraph*{\bf Caveats and systematic uncertainties} The use of the $R_\ell$ ratio was historically motivated due to the cancellation of several uncertainties.  
At \fccee/CEPC it is an open question how to best utilize the large improvement in the statistical precision for $Z\to \ell\bar\ell$.  
The current world average, $R_\ell = 20.767 \pm 0.025$, has about a $10^{-4}$ relative uncertainty, whereas the relative uncertainty of ${\cal B}(Z\to \ell^+\ell^-) = (3.3658 \pm 0.0023)\,\%$ is nearly an order of magnitude larger~\cite{PDG24}.
Normalizing to $\Gamma(Z\to \mbox{hadrons})$ eliminates several sources of uncertainties, but introduces a dependence on $\alpha_s$, since $\Gamma(Z\to \mbox{hadrons}) \propto 1+ \alpha_s/\pi + {\cal O}(\alpha_s^2)$.
At present, the world average of $\alpha_s(m_Z)=0.1180 \pm 0.0009$~\cite{PDG24} has a nearly 1\% relative uncertainty, which is expected to be reducible to $10^{-3}$~\cite{dEnterria:2022hzv}.  

The projected experimental systematic (statistical) uncertainty of the $R_\ell$ measurement is $\delta R_\ell \approx 2\times 10^{-4}$ ($6\times 10^{-5}$), amounting to $10^{-5}$ ($3\times10^{-6}$) relative uncertainty~\cite{FCC:2018byv}. 
Since $\delta R_\ell / R_\ell \approx (\delta \alpha_s/\alpha_s)(\alpha_s/\pi)$,
achieving a comparable theory uncertainty would require a determination of $\alpha_s$ with about $3\times 10^{-4}$ ($1\times 10^{-4}$) relative precision.  
This is somewhat beyond current projections, although unexpected theoretical breakthroughs may well be made by the time \fccee operates.

Alternatively, one may choose to constrain \mbox{$\Gamma(Z\to \ell\bar\ell)$} directly, which would require exquisite precision on the luminosity and detector acceptances.  
Small angle Bhabha scattering promises a relative luminosity uncertainty around $10^{-4}$; furthermore, it may be possible to measure the integrated luminosity even better, with a few times $10^{-5}$ relative uncertainty~\cite{CarloniCalame:2019dom, blondel_2023_f1fs5-0jr59, Dam:2021sdj, JanotLumi}.
A similarly accurate determination of the electromagnetic fine structure constant $\alpha(m_Z)$ would also be needed in this case. 

The sensitivity of $Z\to \ell\bar\ell$ to new physics motivates all possible approaches to reduce the systematic uncertainty of its interpretation, ideally to the statistically achievable precision of this measurement.

\medskip
\paragraph*{\bf Summary \& discussion} 
The measurements at the $Z$ pole, $Zh$, $WW$ and $t\bar t$ thresholds can be combined and parametrized in terms of the Standard Model effective theory (SMEFT).  
Because of its generality, however, the SMEFT has too many parameters to allow for an unambiguous interpretation of the data. 
This is typically addressed by setting all but one or two SMEFT parameters to zero, in a somewhat ad hoc matter.
Concrete (simplified) models on the other hand predict nontrivial correlations between SMEFT operators and allow for interpretations in terms of theoretical puzzles, such as the hierarchy problem, dark matter, etc. 

Our examples indicate that it is important to avoid oversimplifying the models: Figure~\ref{fig:delta Rl Bino} shows that the correction to $R_\ell$ rapidly decreases as either the $\Wino$ or $\slepL$ mass is taken to be very heavy.
This is because the dominant contribution comes from the non-universal vertex correction, for which both particles are important (see Appendix~\ref{App:Details of SUSY Corrections to $Z$ Decay Rate}). 
These vertex corrections also affect other observables, such as effective Weinberg angles ($\sin \theta_{\rm eff}^f$) for various fermion flavors. 
Flavor non-universality is moreover a powerful tool if the sfermion masses are not flavor universal themselves. 
In this case, one may expect deviations in, e.g.,~\mbox{$\Gamma(Z\to \mu^+\mu^-)/\Gamma(Z\to e^+e^-)$}, which may be better constrained than $R_\ell$.

Our analysis also highlights the critical importance of multiple independent measurements of Standard Model parameters, in particular $\alpha_s(m_Z)$.  
For the $R_\ell$ measurement to effectively constrain new physics, an independent determination of $\alpha_s(m_Z)$ with comparable or greater precision is essential. 
If this can be achieved, our results show that the $Z$ factories can probe TeV scale electroweakinos, which may remain undetected by the HL-LHC. This is possible in models with compressed mass spectra or R-parity violation.

\medskip
\paragraph*{\bf Acknowledgements} 

We thank Maurice Garcia-Sciveres, Christophe Grojean, Carl Haber, Lawrence Hall, Timon Heim, Gudrun Hiller, Patrick Janot, Michelangelo Mangano, Matthew McCullough, Simone Pagan Griso, Michael Peskin, Dean Robinson, Martin Schmaltz, Marjorie Shapiro and Elliot Lipeles for helpful conversations. 
We thank Christophe Grojean and Quentin Bonnefoy for comments on the manuscript. 
ZL and SK thank the Aspen Center for Physics (supported by the NSF Grant PHY-1607611) for hospitality while this work was completed.
This work is supported in part by the Office of High Energy Physics of the U.S.\ Department of Energy under contract DE-AC02-05CH11231. 


\bibliographystyle{apsrev4-1}

\bibliography{references.bib}

\clearpage
\onecolumngrid
\newpage

\begin{center}
   \textbf{\large SUPPLEMENTARY MATERIAL \\[.2cm] 
   ``Imprints of supersymmetry at a future $Z$ factory''}\\[.3cm]
  {Simon Knapen, Kevin Langhoff, and Zoltan Ligeti}
\end{center}

\setcounter{equation}{0}
\setcounter{figure}{0}
\setcounter{table}{0}
\setcounter{section}{0}
\setcounter{page}{1}
\makeatletter

\onecolumngrid

\section{Simplified SUSY Models} \label{App:MSSM}

In this appendix, we define the simplified models we use to characterize the effects of particles in the Minimal Supersymmetric Standard Model (MSSM) on $Z$-boson decays. These models are defined using only two mass parameters and the mixing angle between the two Higgs doublets' vevs in the MSSM. Concretely with $H_u$ and $H_d$, the up and down-type Higgs fields in the MSSM, we use the conventional definition $\tan\beta\equiv \langle H_u\rangle/\langle H_d\rangle$. For further details on the MSSM, see Ref.~\cite{Martin:1997ns}.

\subsection{The \texorpdfstring{$\Bino+\slepR$}{} simplified model}

The first simplified model consists of a pure bino $(\Bino)$ and a right-handed (RH) selectron $(\slepR)$, so we include the following interaction terms in the Lagrangian:
\begin{align}\label{eq:EBdefinition}
    \mathcal{L}_{ {\rm int}} \supset ig' B^\mu \slepR^* \overset{\leftrightarrow}{\partial}_\mu \slepR + g'^2 B_\mu B^\mu |\slepR|^2 -\sqrt{2}g'\left(\slepR^* E\Bino + {\rm h.c.}\right)  -\frac{g'^2 c_{2\beta}}{2}|\slepR|^2|H|^2 \,,
\end{align}
where $E$ and $B_\mu$ are the standard model RH electron and $U(1)_Y$ gauge boson respectively and $c_{2\beta}\equiv\cos2\beta$. 
We work in the alignment limit, such that the SM Higgs doublet corresponds to the combination
\begin{align}
    H = \sin \beta H_u + \cos \beta H_d^c \,,
\end{align}
and assume that the four other Higgs sector mass eigenstates are heavy enough to not contribute to $R_\ell$. The fourth term in \eqref{eq:EBdefinition} arises from the hypercharge $D$-term, $\mathcal{L}\supset -(g'^2/2)D_Y^2$ where $D_Y = \sum_{\phi} \phi^* Y \phi$, where the sum goes over all scalars charged under $U(1)_Y$.

\subsection{The \texorpdfstring{$\Wino+\slepL$}{} simplified model}

The second simplified model consists of a pure wino $(\Wino)$ and a left-handed slepton $(\slepL)$ with interaction terms
\begin{align}
    \mathcal{L}_{ {\rm int}} &\supset i \slepL^*\left(g W^{a\mu}\tau^a - \frac{g'}{2}B^\mu \right) \overset{\leftrightarrow}{\partial}_\mu \slepL + \left(\frac{g^2}{4} W^a_{\mu} W^{a\mu}+\frac{{g'}^2}{4} B_{\mu} B^{\mu} \right)|\slepL|^2 \nonumber\\
    &- \sqrt{2}g\left(\slepL^* \tau^a L\Wino^a + {\rm h.c.}\right)+\frac{g'^2 c_{2\beta}}{4} |\slepL|^2|H|^2 - c_{2\beta}g^2 (\slepL^\dagger \tau^a \slepL)(H^\dagger \tau^a H) \,.\label{eq:LWdefinition}
\end{align}
Here $\tau^a$ are the $SU(2)$ generators normalized to $\Tr(\tau^a \tau^b) = \frac{1}{2}\delta^{ab}$. In the broken phase, the last term in \eqref{eq:LWdefinition} causes a slight mass splitting between the sneutrino and selectron component of the $\slepL$ doublet. 
This term is responsible for the non-vanishing $S$ and $T$ parameters, but the vertex correction to $Z\to \ell\ell$ is insensitive to it to leading order in $m_Z/m_{\mathrm{SUSY}}$. It can play a role in LHC phenomenology, since the slight splitting between the neutral and charged component of $\slepL$ implies that the sneutrino can be the LSP.

\section{Details of SUSY Corrections to \texorpdfstring{$Z$}{} Decay Rates}
\label{App:Details of SUSY Corrections to $Z$ Decay Rate}

\subsection{General Structure of SUSY Loop Corrections}

In this section we investigate the SUSY corrections to the ratio of decay rates
\begin{align}
    R_{\ell} = \frac{{\rm BR}(Z \to {\rm hadrons})}{{\rm BR}(Z \to \bar{\ell} \ell)} \,.
\end{align}
The decay rate is given by 
\begin{align}
    \Gamma(Z\rightarrow f \bar{f}) = \frac{1}{3}\frac{1}{16\pi m_Z} \sum \big|\mathcal{M}(Z\to f \bar{f})\big|^2.
\end{align}
The SM amplitude is given by 
\begin{align} \label{eq:M_SM}
    \mathcal{M}_{\rm SM}(Z\rightarrow f \bar{f}) = -\frac{i g}{\hat c}  \epsilon_\mu(Z) \bar{u}(f) \gamma^\mu\left(g^f_L P_L+g^f_R P_R\right) v(\bar{f}) + \mathcal{O}({\rm SM~loops}),
\end{align}
where 
\begin{align}
    g^{f}_L = I_3^f - \hat s^2 Q^f ,\qquad
    g^{f}_R = -\hat s^2Q^f,
\end{align}
$P_{L,R} = (1\mp\gamma_5)/2$, $I_3^f$ and $Q^f$ are respectively the weak and electromagnetic charges of $f$. $\hat s$ is the renormalized value for the sine of Weinberg angle; we specify our choice of scheme in the next section. We further define $\hat c \equiv\sqrt{1-\hat s^2}$. $g$ is the $SU(2)$ gauge coupling.  
Therefore, at tree level in the SM we have
\begin{align}
    \Gamma_{\rm SM}(Z\rightarrow f \bar{f}) = N^f_c\, \frac{\alpha_W m_Z}{6  c_W^2 } \left[(g_L^f)^2+(g_R^f)^2\right] ,
\end{align}
with $N_c^f=3$ ($N_c^f=1$) for quark (lepton) final states and $\alpha_W\equiv g^2/4\pi$.

If we write $\mathcal{M}(Z\rightarrow f \bar{f}) = \mathcal{M}_{\rm SM}(Z\rightarrow f \bar{f}) + \mathcal{M}_{\rm SUSY}(Z\rightarrow f \bar{f})$, where $\mathcal{M}_{\rm SUSY}$ is the 1-loop amplitude induced by the SUSY sector, then the leading corrections to the decay rate are
\begin{align}
    \delta \Gamma(Z\rightarrow f \bar{f}) = \frac{1}{3}\frac{1}{8\pi m_Z}\, {\rm Re}\left[\mathcal{M}_{\rm SM}(Z\rightarrow f \bar{f})\, \mathcal{M}_{\rm SUSY}^*(Z\rightarrow f \bar{f})\right].
\end{align}
The amplitude from one-loop SUSY corrections can be written (following Ref.~\cite{Illana:2002tg}) as
\begin{align}
    \mathcal{M}_{\rm SUSY}=-i  \frac{g^3}{16 \pi^2 c_W} \varepsilon_\mu(Z) \bar{u}\left(f\right)\left[\gamma^\mu\left(f^f_L P_L+f^f_R  P_R \right)+\frac{\sigma_{\mu \nu} Z^\nu}{M_W}\left(i f^f_M+f^f_E \gamma_5\right)\right] v\left(\bar{f}\right).
\end{align}
Using this expression, we can write the SUSY correction to the decay rate as
\begin{align}
    \delta \Gamma(Z\rightarrow f \bar{f}) \approx 
 N^f_c\, \frac{ \alpha_W^2 m_Z }{12 \pi  c_W^2}\left(f^f_L g^f_L+f^f_R g^f_R\right)\implies \frac{\delta \Gamma(Z\rightarrow f \bar{f})}{ \Gamma_{\rm SM}(Z\rightarrow f \bar{f})} = \frac{\alpha_W}{2\pi}\times  \frac{f^f_L g^f_L+f^f_R g^f_R}{(g^f_L)^2+ (g^f_R)^2} \,,\label{eq:susyamplitude}
\end{align}
where we see that the magnetic and electric dipole terms in \eqref{eq:susyamplitude} do not contribute to the interference with the SM.

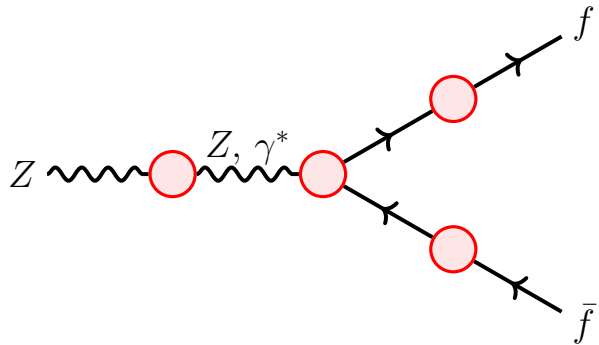
\begin{figure}[tb]
         \begin{tikzpicture}[line width=1.5 pt, scale=1.0]
             \tikzset{blob/.style={circle, draw=red!100, fill=red!10, very thick, minimum size=.6cm}}
            \node (Z) at (-1,0) {\Large $Z$};
            \node at (2,.35) {\Large $Z,\, \gamma^*$};
            \node[blob] (a) at (1,0) {};
            \node[blob] (b) at (3,0) {};
            \draw[vector] (Z)--(a); 
            \draw[vector] (a)--(b); 
            \begin{scope}[xshift = 3cm]
                 \node[blob] (ctop) at (30:2) {};
                 \node[blob] (cbot) at (-30:2) {};
                 \node (dtop) at (30:4) {\Large $f$};
                 \node (dbot) at (-30:4) {\Large $\bar{f}$};
                 \draw[fermion] (b)--(ctop); 
                 \draw[fermionbar] (b)--(cbot); 
                 \draw[fermion] (ctop)--(dtop); 
                 \draw[fermionbar] (cbot)--(dbot);
            \end{scope}
         \end{tikzpicture}
     \caption{General form of $Z\rightarrow \bar{f} f$ diagrams. The vacuum polarization contributions for the gauge boson are captured in the correction to the Weinberg angle $\delta s_\theta$; the vertex correction  and fermion renormalization terms make up the $\tilde f^f_{L,R}$ in Eq.~\eqref{eq:defftilde}.\label{fig:feyn}}
 \end{figure}

The coefficients $f_{L,R}^f$ receive finite direct contributions from vertex corrections and the self energies of the final state fermions which we denote by $\tilde{f}_{L,R}^f$ (see Fig.~\ref{fig:feyn}). 
There is also an indirect contribution from the $\gamma$--$Z$ gauge boson vacuum polarization corrections, which modifies the Weinberg angle in the interactions with the $Z$ boson by a finite value $\delta s^2$ relative to the renormalized value $\hat s^2$ (i.e., $s_{\rm eff}^2 = \hat s^2 +\delta s^2$) as explained in greater detail below. We can therefore write
\begin{equation}\label{eq:defftilde}
f_{L,R}^f=\tilde f_{L,R}^f - \delta s^2\, Q^f. 
\end{equation}
We provide the expression for $\delta s^2$ in terms of the oblique parameters in Eq.~(\ref{eq:rendeltaS}).

In terms of these quantities, the modification to the ratio $R_\ell$ is given by 
\begin{align} \label{eq:R_l}
    \frac{\delta R_\ell}{R_\ell} &=  
    \frac{\delta \Gamma(Z\rightarrow {\rm hadrons} ) }{\Gamma(Z\rightarrow {\rm hadrons})} - \frac{\delta \Gamma(Z\rightarrow \ell \bar{\ell})}{\Gamma(Z\rightarrow \ell \bar{\ell}  )}\notag 
    \\
    &=
    -\frac{\alpha_W}{2\pi}\times  \frac{\tilde{f}^\ell_L (\hat{s}^2-1/2 )+\tilde{f}^\ell_R \hat{s}^2}{(\hat{s}^2-1/2 )^2+ \hat{s}^4} +  \frac{32 (2 \hat{s}^2-3) (5 \hat{s}^2-1)}{\left(8 \hat{s}^4-4 \hat{s}^2+1\right) \left(88 \hat{s}^4-84 \hat{s}^2+45\right)}\, \delta s^2\notag\\
      &\approx
    0.012\times\tilde{f}^\ell_L - 0.0099\times\tilde{f}^\ell_R -0.80\times\delta s^2.
\end{align}
In the second line we dropped vertex corrections to the decay to hadrons, corresponding to taking $\tilde f_{L,R}^q=0$, as they are expected to be negligible for the squark and gluino masses not already excluded by the LHC. 
We verify this point in section~\ref{App:G+Q} (see Figure~\ref{fig:g+q}).

\subsection{Corrections to Weinberg Angle} \label{sec: Corrections to Weinberg Angle}

In this section we give the general expression how the effective Weinberg angle in Eq.~(\ref{eq:M_SM}) is modified by vacuum polarization contributions from SUSY particles (this discussion partially follows Ref.~\cite{Schwartz:2014sze}).
We choose to define the Weinberg angle through the measurements of $\hat\alpha(m_Z)$, $\hat m_Z$ and $\hat G_F$
\begin{equation}
    \hat s^2\, \hat c^2\equiv\frac{\pi \hat\alpha(m_Z)}{\sqrt{2} \hat G_F \hat m_Z^2} \,,\label{eq:schemechoice}
\end{equation}
where hatted $(~\hat{}~)$ quantities refer to the experimentally measured value of the parameter in question (see Table~\ref{tab:params}). 
We define $\hat s\equiv \sin\hat\theta_W$ and $\hat c^2\equiv1-\hat s^2$, with $\hat \theta_W$ the renormalized Weinberg angle. 

\begin{table}[b]
\centering
    \def\arraystretch{1.4}\tabcolsep 10pt
    \begin{tabular}{lll}
    \hline\hline
       Parameter  & Value & Ref.  \\
       \midrule
       $\hat{G}_F$ & $1.1663788(6) \times 10^{-5} ~\mathrm{GeV}^{-2}$ & \cite{MuLan:2010shf,PDG24} \\
       $\hat{m}_Z$ & 91.1876(21) & \cite{ALEPH:2005ab,PDG24} \\
       $\hat{\alpha}(m_Z)$ &  0.02783(6) & \cite{Fanchiotti:1992tu,PDG24}\\
    \hline\hline
    \end{tabular}
    \caption{Experimentally measured input parameters.\label{tab:params}}
\end{table}

The corresponding unrenormalized (bare) parameters are denoted by $s,c$. 
This scheme choice is motivated by excellent experimental precision on $\hat\alpha(m_Z)$, $\hat m_Z$ and $\hat G_F$, and the fact that the IR determinations of $\hat\alpha(m_Z)$ and $\hat G_F$ in particular are insensitive to new physics above the $Z$-pole. These experimental inputs are related to the bare Lagrangian parameters (without~$\hat{}$~) by
\begin{align}
    \hat \alpha(m_Z)&= \alpha \left(1+\frac{\Pi_{\gamma\gamma}(m_Z^2)}{m_Z^2}\right) , \label{eq:renalpha}\\
    \hat m_Z^2 &= m_Z^2 + \mathrm{Re}\left[\Pi_{ZZ}(m_Z^2)\right] , \label{eq:renmZ}\\
    \hat G_F &= G_F \left(1-\frac{\Pi_{WW}(0)}{m_W^2}\right), \label{eq:renGF}
\end{align}
where the $\Pi_{VV}$ are the contributions to the vacuum polarizations from the supersymmetric particles. The bare Weinberg angle satisfies the same relation with $\alpha(m_Z)$, $m_Z$ and $G_F$ as for the renormalized parameters in \eqref{eq:schemechoice} and is related to the measure quantities through the relation
\begin{align} \label{eq:s^2}
    s^2=\hat{s}^2+\frac{\hat{s}^2\hat{c}^2}{\hat{c}^2-\hat{s}^2}\left(-\frac{\Pi_{\gamma \gamma}\left(\hat{m}_Z^2\right)}{\hat{m}_Z^2}+\frac{\Pi_{Z Z}\left(\hat{m}_Z^2\right)}{\hat{m}_Z^2}-\frac{\Pi_{W W}(0)}{\hat{m}_W^2}\right) .
\end{align}
Further, the effective Weinberg angle in the $Z\bar f f$ vertex is related to the bare Weinberg angle through the relation 
\begin{align}
    s_{\rm{eff}}^2 = s^2-s c\, \frac{\Pi_{\gamma Z}\left(m_Z^2\right)}{m_Z^2} \,.
\end{align}
Replacing the bare value of $s^2$ in the above equation using (\ref{eq:s^2}) results in the following finite difference for the two definitions of the Weinberg angle
\begin{align}
\delta s^2 \equiv s_{\rm eff}^2 - \hat{s}^2 
=\frac{\hat{s}^2 \hat{c}^2}{\hat{c}^2-\hat{s}^2}
\left(-\frac{\Pi_{\gamma \gamma}\left(\hat{m}_Z^2\right)}{\hat{m}_Z^2}+\frac{\Pi_{Z Z}\left(\hat{m}_Z^2\right)}{\hat{m}_Z^2}-\frac{\Pi_{W W}(0)}{\hat{m}_W^2}\right)
-\hat{s} \hat{c}\, \frac{\Pi_{\gamma Z}\left(\hat{m}_Z^2\right)} {\hat{m}_Z^2} \,,
\label{eq:rendeltaS}
\end{align}
where in the second line we used \eqref{eq:renalpha}, \eqref{eq:renmZ}, \eqref{eq:renGF} and the unhatted version of \eqref{eq:schemechoice} to substitute the bare parameters $s,c$ for their renormalized values $\hat s, \hat c$, such that \eqref{eq:rendeltaS} is free of logarithmic divergences. 
It can further be rewritten in terms of the Peskin-Takeuchi parameters~\cite{Peskin:1991sw,Barbieri:2004qk} defined as
\begin{align}
    S &\equiv -\frac{4s^2}{\alpha}\frac{g}{g^{\prime}} \Pi_{W_3 B}^{\prime}(0)\,,\label{eq:defS}\\
    T &\equiv -\frac{1}{\alpha}\frac{\Pi_{W_3 W_3}(0)-\Pi_{W^{+} W^{-}}(0)}{M_W^2}\,.\label{eq:defT}
\end{align}
This gives
\begin{align}
    \delta s^2 = \frac{\alpha}{\hat c^2 - \hat s^2}\left[\frac{1}{4}S - \hat s^2\hat c^2 T\right] .
\end{align}
It now remains to calculate the $\tilde f_{L,R}^f$, $S$ and $T$ for the models of interest which we do in \ref{App:B+E} and \ref{App:W+L}. This is most easily done in terms of the Veltman-Passarino functions which are reviewed in \ref{App:Passarino-Veltman Functions}. 

\subsection{Passarino-Veltman Functions}
\label{App:Passarino-Veltman Functions}

It is convenient to use the Veltman-Passarino functions \cite{Passarino:1978jh,Denner:1991kt} to describe the one-loop corrections to the vertex factors from SUSY particles which we define in this section. These functions are of the form 
\begin{align}
    \mathbf{A}_0(m) &= \left(\frac{i}{16\pi^2}\right)^{-1}\mu^{d-4}\int \frac{d^d\ell}{(2\pi)^d} \frac{1}{[\ell^2 - m^2]} \,,
    \\
    \mathbf{B}_{\mu_1,..., \mu_k}(p,m_1,m_2) &= \left(\frac{i}{16\pi^2}\right)^{-1}\mu^{d-4}\int \frac{d^d\ell}{(2\pi)^d} \frac{\ell_{\mu_1}\cdots \ell_{\mu_k}}{[\ell^2 - m_1^2  ][(\ell+p)^2 - m_2^2]} \,,
    \\
    \mathbf{C}_{\mu_1,..., \mu_k}(p_1,p_2,m_1,m_2,m_3) &= \left(\frac{i}{16\pi^2}\right)^{-1}\mu^{d-4}\int \frac{d^d\ell}{(2\pi)^d} \frac{\ell_{\mu_1}\cdots \ell_{\mu_k}}{[\ell^2 - m_1^2   ][(\ell+p_1)^2 - m_2^2   ][(\ell+p_2)^2 - m_3^2]} \,.
\end{align}

The scalar integrals where the numerator is given by $1$ are denoted as $\mathbf{B}_0$ and $\mathbf{C}_0$. The tensor integrals can be decomposed as
\begin{align}
    \mathbf{B}_\mu &= p_\mu \mathbf{B}_1 \,,
    \\
    \mathbf{B}_{\mu \nu} &= g_{\mu \nu} \mathbf{B}_{00} + p_{\mu}p_{\nu} \mathbf{B}_{11} \,,
    \\
    \mathbf{C}_\mu &= p^1_\mu \mathbf{C}_1 + p^2_\mu \mathbf{C}_2 \,,
    \\
    \mathbf{C}_{\mu\nu} &= g_{\mu \nu} \mathbf{C}_{00} + p^1_\mu p^1_\nu \mathbf{C}_{11}+ p^2_\mu p^2_\nu \mathbf{C}_{22}+ p^1_\mu p^2_\nu \mathbf{C}_{12} \,.
\end{align}
These functions only depend on the Lorentz invariant inner products of four momenta, and we use the convention for the order of the arguments $\mathbf{B}_{\cdots}(p^2,m_1,m_2)$ and $\mathbf{C}_{\cdots}(p_1^2,(p_1+p_2)^2,p_2^2,m_1,m_2,m_3)$.

\subsection{The \texorpdfstring{$\Bino + \slepR$}{} simplified model} \label{App:B+E}

The first scenario we consider is the pure bino scenario where we consider the two particles $\Bino$ and $\slepR$ with masses $M_{\Bino}$ and $M_{\slepR}$. Since the $\slepR$ interacts exclusively with the right-handed SM leptons one finds $\tilde{f}_{L} = 0$. For $\tilde{f}_{R}$, there exists only two different types of diagrams which contribute as shown in Fig.~\ref{fig:Bino_Diagrams}. These give (in order of left to right) the following contributions to the coefficient $f_R$:
\begin{align}
    \tilde{f}_{R,\,1} &= \frac{4s_W^4}{c_W^2} \mathbf{C}_{00}\left(m_Z^2,0,  0, M_{\Bino}^2, M_{\slepR}^2, M_{\slepR}^2\right) ,
    \\
    \tilde{f}_{R,\,2} &= \frac{2s_W^4}{c_W^2} \mathbf{B}_1\left(0, M_{\Bino}, M_{\slepR}\right) .
\end{align}
To leading order in small $m_Z$, this is approximated by the following:
\begin{align}
    \tilde{f}_{R} = \tilde{f}_{R,\,1} +\tilde{f}_{R,\,2} \approx \frac{s_W^4 m_Z^2}{18c_W^2}\left[
    \frac{2 M_{\slepR}^6-9 M_{\Bino}^2 M_{\slepR}^4+18 M_{\Bino}^4 M_{\slepR}^2 -11 M_{\Bino}^6 +6 M_{\Bino}^6 \ln \big({M_{\Bino}^2}/{M_{\slepR}^2}\big)}{ \big(M_{\Bino}^2-M_{\slepR}^2\big)^4}
   \right] + \mathcal{O}\left( \frac{m_Z^4}{M_{\Bino,\slepR}^4}\right) .
\end{align}
The $\slepR$ is charged exclusively under hypercharge which implies
\begin{equation}
S=T=0 \,.
\end{equation}

\begin{figure}[tb]
    \centering
    \begin{minipage}{0.33\textwidth}
        \centering
        \begin{tikzpicture}[line width=1.5 pt, scale=1.4]
            \draw[vector] (1,0)--(2,0);
            \draw[scalarnoarrow,red] (2,0)--(3,.5);
            \draw[scalarnoarrow,red] (2,0)--(3,-.5);
            \draw[fermion,red] (3,-.5)--(3,.5);
            \draw[fermion] (3,.5)--(4,1);
            \draw[fermionbar] (3,-.5)--(4,-1);
            \node[rotate=30] at (2.4,0.5) {\normalsize \textcolor{red}{$\slepR$}};
            \node[rotate=-30] at (2.4,-0.5) {\normalsize \textcolor{red}{$\slepR$}};
            \node at (3.3,0.0) {\normalsize \textcolor{red}{$\Bino$}};
            \node at (0.8,0.) {$Z$};
            \node at (4.25,1.05) {\normalsize $\ell$};
            \node at (4.25,-0.95) {\normalsize $\bar{\ell}$};
        \end{tikzpicture}
    \end{minipage}
    \hspace{.2cm}
    \begin{minipage}{0.3\textwidth}
        \centering
        \begin{tikzpicture}[line width=1.5 pt, scale=1.4]
            \draw[vector] (1,0)--(2,0);
            \draw[fermion] (2,0)--(2.666,.3333);
            \draw[fermion,red] (2.666,.3333)--(3.333,.6666);
            \draw[fermion] (3.333,.6666) -- (4,1);
            \draw[fermionbar] (2,0)--(4,-1);
            \draw[scalarnoarrow,red] (2.666,.3333) to[out= -60, in = -60,looseness = 2] (3.333,.6666);
            \node[rotate=0] at (3.6,0.2) {\normalsize \textcolor{red}{$\slepR$}};
            \node[rotate=0] at (3.,0.8) {\normalsize \textcolor{red}{$\Bino$}};
            \node at (0.8,0.) {$Z$};
            \node at (4.25,1.05) {\normalsize $\ell$};
            \node at (4.25,-0.95) {\normalsize $\bar{\ell}$};
        \end{tikzpicture}
    \end{minipage}
    \caption{The two types of diagrams contributing to the SUSY correction to the $Z$ boson decay rate when considering only the contributions from a pure bino and the right-handed sleptons. 
    }
    \label{fig:Bino_Diagrams}
\end{figure}

\subsection{The \texorpdfstring{$\Wino + \slepL$}{} simplified model}\label{App:W+L}

The next scenario we consider is the pure wino scenario where we consider the particles $\Wino^{\pm,0}$, $\slepL$, and $\widetilde{\nu}_\ell$, where all the winos are assumed to have degenerate mass, $M_{\Wino}$, and the left-handed sleptons and sneutrinos have degenerate mass, $M_{\slepL}$. 
There exists two different types of diagrams which contribute as shown in Fig.~\ref{fig:Wino_Diagrams}. These give the (in order of left to right) the following contributions to the coefficient $\tilde{f}_L$:
\begin{align}
    \tilde{f}_{L,\, 1} &=\frac{3-2c_W^2}{2}\, \mathbf{C}_{00}\left(m_Z^2,0,  0, M_{\Wino}^2, M_{\slepL}^2, M_{\slepL}^2\right) ,
    \\
    \tilde{f}_{L,\, 2} &= \frac{3-6c_W^2}{4}\, \mathbf{B}_1\left(0, M_{\Wino}, M_{\slepL}\right) ,
    \\
    \tilde{f}_{L,\, 3} & = c_W^2 \bigg[ M_{\Wino}^2\, \mathbf{C}_{0}\left(m_Z^2,0,  0,  M_{\slepL}^2, M_{\Wino}^2,M_{\Wino}^2\right) + m_Z^2\,
    \mathbf{C}_{12}\left(m_Z^2,0,  0,  M_{\slepL}^2, M_{\Wino}^2,M_{\Wino}^2\right) \notag
    \\
    & \qquad -
    2\, \mathbf{C}_{00}\left(m_Z^2,0,  0,  M_{\slepL}^2, M_{\Wino}^2,M_{\Wino}^2\right) + \frac{1}{2}
    \bigg],\\
\tilde{f}_L &=\sum_{i=1}^3 \tilde{f}_{L,i} \,.
\end{align}

To leading order in $m_Z$, this can be approximated as
\begin{align}
    \tilde{f}_L = \frac{m_Z^2}{48 \big(M_{\slepL}^2-M_{\Wino}^2\big)^4} \bigg[& 
    2\left(1+10 c_W^2\right) M_{\slepL}^6-9 \left(6 c_W^2+1\right) M_{\slepL}^4 M_{\Wino}^2+18 \left(2 c_W^2+1\right) M_{\slepL}^2 M_{\Wino}^4-\left(2
   c_W^2+11\right) M_{\Wino}^6 \notag \\
   -&4 \left( 8 c_W^2 M_{\slepL}^6-12 c_W^2 M_{\slepL}^4 M_{\Wino}^2+(3-2c_W^2) M_{\Wino}^6\right) \ln \left(\frac{M_{\slepL}}{M_{\Wino}}\right)
    \bigg] +  \mathcal{O}\bigg( \frac{m_Z^4}{M_{\Wino,\slepL}^4}\bigg) ,
\end{align}
while $\tilde{f}_R=0$. The oblique parameters are \cite{Marandella:2005wc}\footnote{Note that the definitions of $S$ and $T$ in \cite{Marandella:2005wc} differ by a sign relative to our definitions in \eqref{eq:defS} and \eqref{eq:defT} and in the PDG review \cite{PDG24}.}
\begin{align}
  S &= -4s_W^2\alpha^{-1}\times \frac{\alpha_W \cos(2\beta)}{16 \pi}\bigg(\frac{m_W^2}{M_{\slepL}^2}\bigg) \,, \\
  T &= -\alpha^{-1}\times\frac{\alpha_W \cos^2(2\beta)}{16 \pi} \bigg(\frac{m_W^2}{M_{\slepL}^2}\bigg) \,,
\end{align}
 with $\tan\beta = \langle H_u\rangle/\langle H_d\rangle$ the ratio of the vacuum expectation values of the Higgs doublets in MSSM. For our numerical results we assume the large $\tan\beta$ limit, effectively setting $\cos(2\beta)=-1$. 

\begin{figure}[t]
    \centering
    \begin{minipage}{0.31\textwidth}
        \centering
        \begin{tikzpicture}[line width=1.5 pt, scale=1.4]
            \draw[vector] (1,0)--(2,0);
            \draw[scalarnoarrow,red] (2,0)--(3,.5);
            \draw[scalarnoarrow,red] (2,0)--(3,-.5);
            \draw[fermion,red] (3,-.5)--(3,.5);
            \draw[fermion] (3,.5)--(4,1);
            \draw[fermionbar] (3,-.5)--(4,-1);
            \node[rotate=30] at (2.4,0.5) {\normalsize \textcolor{red}{$\slepL$}};
            \node[rotate=-30] at (2.4,-0.5) {\normalsize \textcolor{red}{$\slepL$}};
            \node at (3.4,0.0) {\normalsize \textcolor{red}{$\Wino$}};
            \node at (0.8,0.) {$Z$};
            \node at (4.25,1.05) {\normalsize $\ell$};
            \node at (4.25,-0.95) {\normalsize $\bar{\ell}$};
        \end{tikzpicture}
    \end{minipage}\hspace{.2cm}
    \begin{minipage}{0.31\textwidth}
        \centering
        \begin{tikzpicture}[line width=1.5 pt, scale=1.4]
            \draw[vector] (1,0)--(2,0);
            \draw[fermion] (2,0)--(2.666,.3333);
            \draw[fermion,red] (2.666,.3333)--(3.333,.6666);
            \draw[fermion] (3.333,.6666) -- (4,1);
            \draw[fermionbar] (2,0)--(4,-1);
            \draw[scalarnoarrow,red] (2.666,.3333) to[out= -60, in = -60,looseness = 2] (3.333,.6666);
            \node[rotate=0] at (3.5,0.) {\normalsize \textcolor{red}{$\slepL$}};
            \node[rotate=30] at (2.9,0.8) {\normalsize \textcolor{red}{$\Wino$ }};
            \node at (0.8,0.) {$Z$};
            \node at (4.25,1.05) {\normalsize $\ell$};
            \node at (4.25,-0.95) {\normalsize $\bar{\ell}$};
        \end{tikzpicture}
    \end{minipage}
    \hspace{.2cm}
    \begin{minipage}{0.31\textwidth}
        \centering
        \begin{tikzpicture}[line width=1.5 pt, scale=1.4]
            \draw[vector] (1,0)--(2,0);
            \draw[fermion,red] (2,0)--(3,.5);
            \draw[fermionbar,red] (2,0)--(3,-.5);
            \draw[scalarnoarrow,red] (3,-.5)--(3,.5);
            \draw[fermion] (3,.5)--(4,1);
            \draw[fermionbar] (3,-.5)--(4,-1);
            \node[rotate=30] at (2.4,0.6) {\normalsize \textcolor{red}{$\Wino$}};
            \node[rotate=-30] at (2.4,-0.6) {\normalsize \textcolor{red}{$\Wino$}};
            \node at (3.3,0.0) {\normalsize \textcolor{red}{$\slepL$}};
            \node at (0.8,0.) {$Z$};
            \node at (4.25,1.05) {\normalsize $\ell$};
            \node at (4.25,-0.95) {\normalsize $\bar{\ell}$};
        \end{tikzpicture}
    \end{minipage}
    \caption{The three types of diagrams contributing to the SUSY correction to the $Z$ boson decay rate when considering only the contributions from a pure wino and the right-handed sleptons. 
    }
    \label{fig:Wino_Diagrams}
\end{figure}

\subsection{The \texorpdfstring{$\tilde g + \tilde q$}{} simplified model} \label{App:G+Q}

The final scenario we consider is where we the light particles are the gluino $\tilde g$ and the squarks $\tilde{q}$ (also investigated in Ref.~\cite{Hagiwara:1990st}).
The $Z$ cannot decay to a $t\bar t$ pair, so the stop squark does not contribute to the vertex correction for flavor diagonal squark masses, given that we take the charginos to be decoupled. 
We also assume that the left and right handed squarks have the same mass and all five flavors of squarks contributing in the loops have the same mass.
The diagrams are identical to those of the $\Bino + \slepR$ under the exchange $\{\Bino,\slepR\} \rightarrow \{\tilde{g},\tilde{q}\}$. This yields 
\begin{align}
    \tilde{f}^{u,d}_{L,R} &= \frac{4}{3} 
    g_{L,R}^{u,d}\frac{\alpha_s}{\alpha_W}\left[\mathbf{C}_{00}\left(m_Z^2,0,  0, M_{\tilde g}^2, M_{\tilde q}^2, M_{\tilde q}^2\right) + \frac{1}{2}\mathbf{B}_1\left(0, M_{\tilde g}, M_{\tilde q}\right)\right]  .
\end{align}

The oblique parameters are taken from \cite{Marandella:2005wc},
\begin{align}
  S&=  -4s_W^2\alpha^{-1}\times \frac{\alpha_W \cos(2\beta)}{16 \pi}\bigg(\frac{m_W^2}{M_{\tilde{q}}^2}\bigg) \,, \\
  T&= -\alpha^{-1}\times\frac{\alpha_W \cos^2(2\beta)}{8 \pi} \bigg(\frac{m_W^2}{M_{\tilde{q}}^2}\bigg) \,.
\end{align}
As already mentioned, we assume the large $\tan\beta$ limit, effectively setting $\cos(2\beta)=-1$. 

\begin{figure}[tb]
     \centering
     \includegraphics[scale = 0.6]{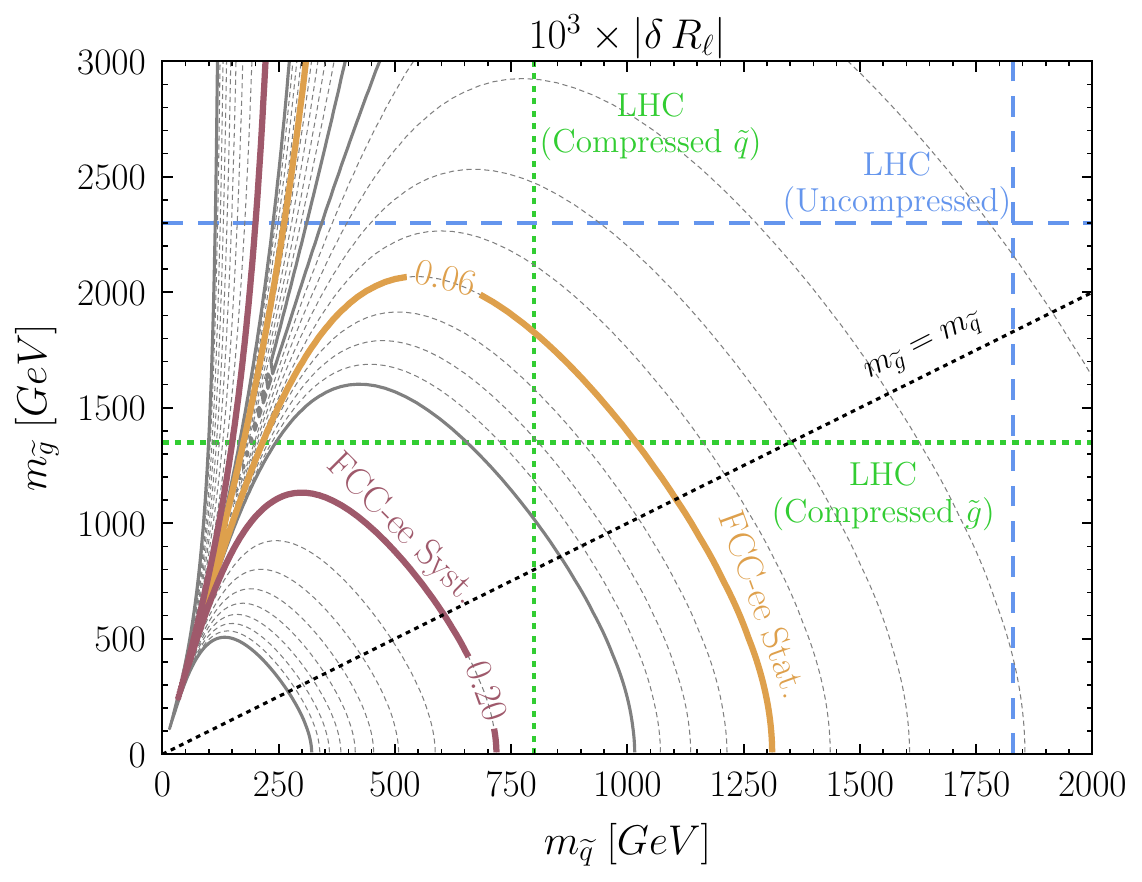}
     \caption{
     $|\delta R^\ell|$ for gluino and squarks simplified model. 
     The \bfYellow{yellow} (\bfRed{red}) curves show the sensitivity for the anticipated \fccee statistical (systematic) uncertainties~\cite{FCC:2018byv}. 
     We also show the current limit from the ATLAS $\tilde{g} \rightarrow q \bar{q} \tilde{\chi}_1^0$ $(\tilde{q} \rightarrow q\tilde{\chi}_1^0)$ search~\cite{ATLAS:2020syg} in the limit $m_{\tilde{\chi}_1^0} \ll m_{\tilde{g}}$ $(m_{\tilde{\chi}_1^0} \ll m_{\tilde{q}})$ as the horizontal (vertical) dashed \bfBlue{blue} line.  The dotted \bfGreen{green} lines show the approximate bounds on the squark and gluino masses if one allows a squeezed spectrum. However, the availability of cascade decays means that these are likely a substantial underestimation of the true bounds in the regime where both $\tilde g$ and $\tilde q$ can be produced at the LHC. A recast of the multi-jet searches is needed to determine this, which we do not attempt in this Letter.}
     \label{fig:g+q}
 \end{figure}

 The funnel region in Figure~\ref{fig:g+q} around $m_{\tilde g}\sim 5 m_{\tilde q}$ is due to destructive interference between the oblique and vertex contributions. We see that the \FCCee will likely not be competitive with the LHC, even when assuming a compressed spectrum and negligible systematic uncertainties for the 
\FCCee. Moreover, the limits on RPV decays of the gluino ($m_{\tilde g}\gtrsim 1800$ GeV \cite{ATLAS:2024kqk}) are already strong enough to comfortably exclude the viable parameter space for the \FCCee, if one assumes the current estimates for the systematic uncertainty on $R_\ell$. 

\section{Details on the interpretation of the LHC limits\label{app:LHClimits}}
In this appendix we elaborate on our reinterpretation of the current LHC bounds in terms of the $\slepL\to \Wino$ cascade decay, with a compressed bino LSP. We use the following notations for the components of the $\slepL$ doublet and $\Wino$ triplet states
\begin{equation}
    \slepL = \left(\begin{array}{c} \tilde \ell_L^+ \\ \tilde \nu\end{array}\right) \quad \mathrm{and} \quad \Wino = \left(\begin{array}{c} \Wino^+ \\ \Wino^0 \\ \Wino^-\end{array}\right),
\end{equation}
with the $\tilde \ell_L^\pm$ the (left-handed) charged slepton and the $\tilde \nu$ the sneutrino. For the $m_{\Wino}<m_{\slepL}$ case with a compressed spectrum we have assumed a bino-like LSP which is compressed relative to the wino NLSP, such that the final states from the $\Wino \to \Bino$ decay are too soft to meaningfully contribute to the limit. At the same time, we have assumed that the splitting is still sufficient for the wino-like NLSP to decay promptly, hence avoiding the  disappearing track signature \cite{ATLAS:2022pib,CMS:2023mny}. A separation of a few GeV between the two neutral states of the wino-bino system suffices to satisfy both conditions. 

To reinterpret the existing $\ell^\pm \ell^\pm$ + MET and $\ell^+\ell^- \,+$~MET searches we need to account for the differences in cross sections and branching ratios relative to those in the benchmark models used by the experimental collaborations. Concretely, the  $\ell^+\ell^-\,+$~MET searches used the $pp \to \tilde \ell_{L,R}^+ \tilde \ell_{L,R}^-$ benchmark model with degenerate left and right-handed sleptons, which decay to $\ell^+\ell^-\,+$~MET with a 100\% branching ratio \cite{ATLAS:2019lff,CMS:2020bfa}. Our simplified model differs in the following ways
\begin{enumerate}
\item The right-handed sleptons are not included and therefore do not contribute to the cross section. 
\item The branching ratio for $\tilde \ell_L^+ \to \ell^+ \Wino^0$ is only 50\%, with the remaining 50\% going to $\tilde \ell_L^+ \to \nu \Wino^+$. Here the $\Wino^\pm$ is assumed to act as missing energy, due to the compressed spectrum discussed above. Similarly the sneutrino can decay to $\tilde \nu\to \ell^+ \Wino^-$, $\tilde \nu\to \nu \Wino^0$ and $\tilde \nu\to \ell^- \Wino^-$, each with a branching ratio of 1/3. 
\item This means that aside from $p p \to \tilde \ell_L^+\tilde \ell_L^-$ production, the $p p \to \tilde \ell_L^+\tilde \nu$, $p p \to \tilde \ell_L^-\tilde \nu$ and $p p \to \tilde \nu\tilde \nu$ channels can also contribute to the $\ell^+\ell^- \,+$~MET rate. We neglect the $p p \to \tilde \ell_L^-\tilde \nu$ channel, as its cross section is substantially smaller than other three.
\end{enumerate}
When accounting for these differences, we find that the effective $\ell^+\ell^- \,+ $~MET cross section in this model is roughly 50\% lower than the $\tilde \ell_{L,R}^+ \tilde \ell_{L,R}^-$ cross section with 100\% branching ratio to $\ell^+\ell^- \,+$~MET. We use this correction factor to rescale the CMS limit in \cite{CMS:2020bfa}, which yields the bound below the dotted line in the right-hand panel of Figure~\ref{fig:delta Rl Bino}. We performed the same procedure on the same sign dilepton search in \cite{CMS:2021cox} and found a bound that is comparable but slightly weaker than that from the $\ell^+\ell^- \,+$~MET search.

\end{document}